# A kilo-Ampere level HTS flux pump

**Jianzhao Geng[1, a)], Tom Painter [2)], Peter Long [1)], Jamie Gawith [1)], Jiabin Yang [1)], Jun Ma [1)], Qihuan Dong[1)], Boyang Shen[1)], Chao Li [1)], and T. A. Coombs [1,a)]**

1) Department of Engineering, University of Cambridge, Cambridge, CB3 0FA, United Kingdom

2) National High Magnetic Field Lab, Tallahassee, FL 32310, USA

a) Email: jg717@cam.ac.uk, tac1000@cam.ac.uk

## Abstract

This paper reports a newly developed high current transformer-rectifier High-$T_c$ Superconducting (HTS) flux pump switched by dynamic resistance. A quasi-persistent current of over 1.1 kA has been achieved at 77 K using the device, which is the highest reported operating current by any HTS flux pumps to date. The size of the device is much smaller than traditional current leads and power supplies at the same current level. Parallel YBCO coated conductors are used in the transformer secondary winding as well as in the superconducting load coil to achieve high current. The output current is limited by the critical current of the load rather than the flux pump itself. Moreover, at over 1 kA current level, the device can maintain high flux injection accuracy, and the overall flux ripple is less than 0.2 mili-Weber. The work has shown the potential of using the device to operate high field HTS magnets in ultra-high quasi-persistent current mode, thus substantially reducing the inductance, size, weight, and cost of high field magnets, making them more accessible. It also indicates that the device is promising for powering HTS NMR/MRI magnets, in which the requirement for magnetic field satiability is demanding.

## 1 Introduction

High-$T_c$ Superconducting (HTS) Coated Conductors (CCs) have shown superior performance in the application of high field magnets [1]. With the increase in the magnetic fields, it is more desirable to use high current cables made of parallel CCs to reduce the inductance of magnets [2]. Compared to the operation model of all conductors connected in series, parallel conductor operation can also increase engineering current density. This is because the critical current of a CC is nonhomogeneous along its length, and it is affected by external magnetic fields. In series operation, the operating current is limited by the portion which has lowest current capacity throughout the whole length. Furthermore, parallel conductor operation makes it less demanding for long length tapes with high critical current, which could reduce the cost of magnets. However, due to joint resistance and flux creep, HTS magnets are normally powered by external power sources via a pair of current leads. The heat load generated by the current leads limits the transport current, impeding high current operations, especially in conduction cooled magnets. Therefore, coils in HTS magnets normally have to be powered in series. Another drawback of using external power supply is the inferior magnetic field stability, which is a main concern for MRI/NMR magnets.

Flux pumps are the kind of devices which can inject flux into a superconducting circuit without electrical contact. They can be used to power closed HTS magnets in quasi-persistent current mode, and could eliminate the drawbacks of using external power supplies. During this decade, several types of HTS flux pumps have been proposed. Hoffmann et al [3] proposed a rotating permanent magnets HTS flux pump (the HTS dynamo). Bai et al [4] proposed a linear travelling wave flux pump. There are also various derivatives and optimizations of these devices [6-14]. In terms of physics, Jiang et al [15] discovered dynamic resistance [16-18] as a limiting factor in the saturation current of the HTS dynamo. Bumby et al [19] attributed the open circuit DC voltage in the HTS dynamo to the rectifying effect due to varying resistances in forward and back screening current paths of the superconducting stator. Geng et al [20] proposed a generalized transformer-rectifier model to explain travelling wave based flux pumps, in which the magnetic fields were considered to play the role of magnetic induction as well as switching, and the switching effect is due to field strength, field rate of change, and current density dependence of flux motion in the superconductors. Based on the understanding of the physics, Geng and Coombs proposed two HTS transformer-rectifier flux pumps. One is switched by dynamic resistance [21], employing field rate of change dependence of flux motion. The other is self-switching [22], which takes advantage of current density dependence of flux motion (flux flow). Recently, Campbell [23] proposed a FEM which confirmed that the field dependence of critical current can result in flux pumping. Geng and Coombs [24] introduced the modeling methodology of field rate of change (dynamic resistance) induced flux pumping.

In terms of operating current, most of the reported results are at the level of 100 A. Recently Hamilton et al [25] reported a squirrel-cage like HTS dynamo design, which has a positive output voltage with a 700 A input current, indicating that the device could possibly output a DC over 1 kA when it is connected to a proper superconducting load. But it would still need experimental proof. In contrast, the dynamic resistance switched transformer-rectifier flux pump has the potential to generate extremely high quasi-persistent current. Due to its clear physics and simple circuit topology, the flux pump has managed to decouple magnetic induction from switching. The structure can substantially reduce loss in superconducting circuit and make quantitative flux injection possible. These merits enable the device to generate high direct current with low ripple. Based on the invention, the Cambridge EPEC Superconductivity Group is collaborating with the National High Magnetic Field Lab (NHMFL) of USA in developing flux-pumped ultra-high current HTS magnets. The preliminary goal is to achieve a flux-pumped HTS solenoid operating at a quasi-persistent current of 5.6 kA at a temperature below 20 K. In this work, we will show the design, development, and test results of a transformer-rectifier flux pump prototype which can output a DC of over 1.1 kA at 77 K. This is the first step of achieving our preliminary goal of 5.6 kA. The results prove that the flux pump is scalable and has the potential to wirelessly power high field HTS magnets at ultra-high current level. Results will also show that the device can maintain a high direct current with extremely low ripple, which is desirable for HTS MRI/NMR magnets.

## 2 The flux pump prototype

### 2.1 the superconducting circuit design

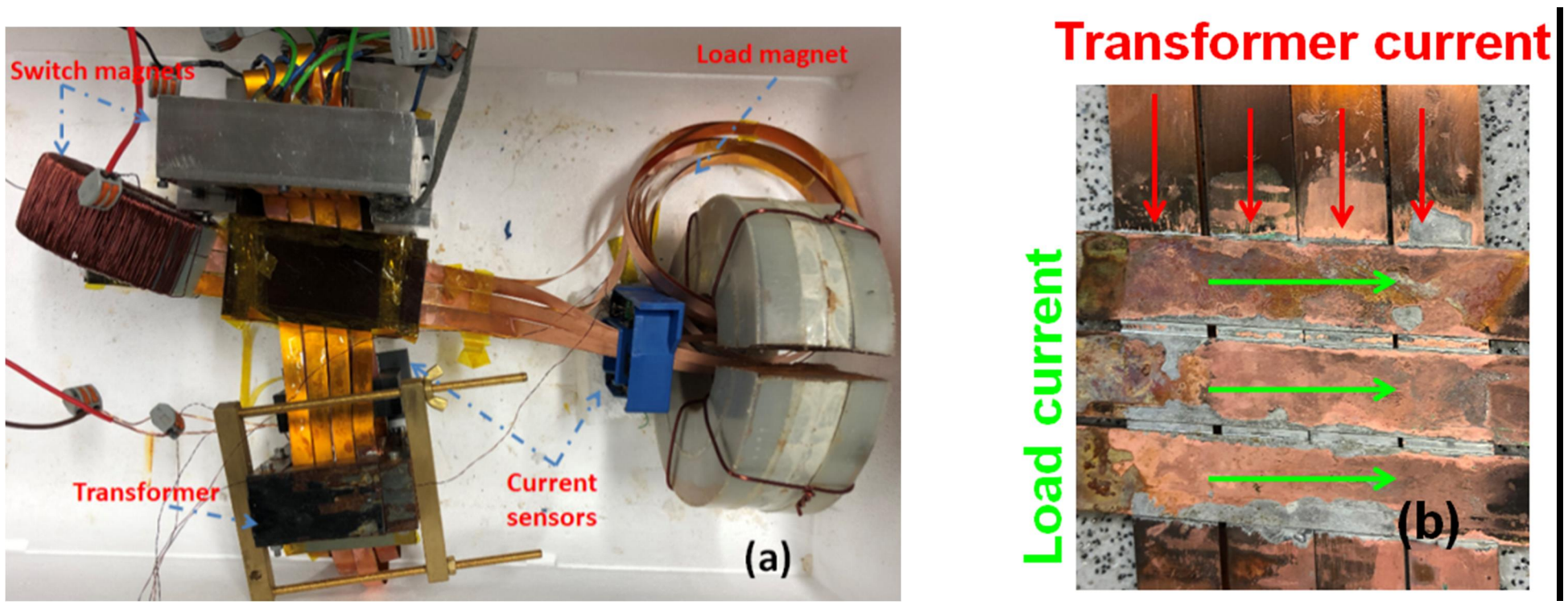

Fig. 1 Pictures of the flux pump circuit. (a) the whole circuit which consists of a transformer, a pair of magnetic switches, a superconducting load magnet, and sensors. (b) the superconducting joints, where four tapes from the transformer and the three tapes from the load are cross-soldered.

The flux pump circuit is shown in Fig. 1. The circuit consists of a transformer, a pair of magnetic switches, a superconducting load magnet, and current sensors. The transformer primary winding is 150 turns of copper, and the secondary is one turn of four parallel superconducting tapes. The load magnet is formed by three parallel superconducting coils linking a gapped large steel core. This is to increase the inductance of the load. Each superconducting coil only has 4 turns. The terminations of each coil were soldered together. The four parallel tapes from the transformer secondary and the three tapes from the load coils' leads are cross-soldered together, as shown in Fig. 1(b). There are 12 (3×4) joints on each side, and 24 (12×2) joints in total. The forward path and backward path of the transformer secondary were closely aligned together, so that the mutual inductance between the secondary loop and the load loop is minimized. The magnetic fields generated by the forward current and the backward current cancels each other, avoiding the $I_c$ reduction in the load terminations due to secondary current generated field. At the portion of tapes which is under the switching AC field, bifilar structure is used, so that the induction generated by the switch magnets is also minimized. All these designs are aimed to reduce the ripple in the load current to achieve a better load field stability, as well as to increase the load current capacity. All tapes used in the superconducting circuit are 12 mm wide and are from *Sunam*. The critical current of the superconducting tapes is labeled to be over 700 A @ 77 K, but limited by our current supply we did not manage to measure the exact value. To acquire the inductance of the load coil, we wound 4 turns of insulated copper wire round the iron core at room temperature. Its inductance was measured by an inductance meter, and the results varied in the range of 7-9 $\mu$H in different measurements. Although there are errors of using the copper winding magnet to estimate the inductance of the superconducting magnet, we assume that the error is within a reasonably low range. The switch consists of two electromagnets with gapped iron cores. One is from a toroid transformer, and the other is from a linear flux pump which has been reported in Ref. [26]. The effective length of the switching field is about 3 cm.

### 2.2 the measurement system and power supplies

The secondary current and the load current are measured by open loop Hall effect current sensors. A picture of the two sensors and their calibrations is shown in Fig. 2.

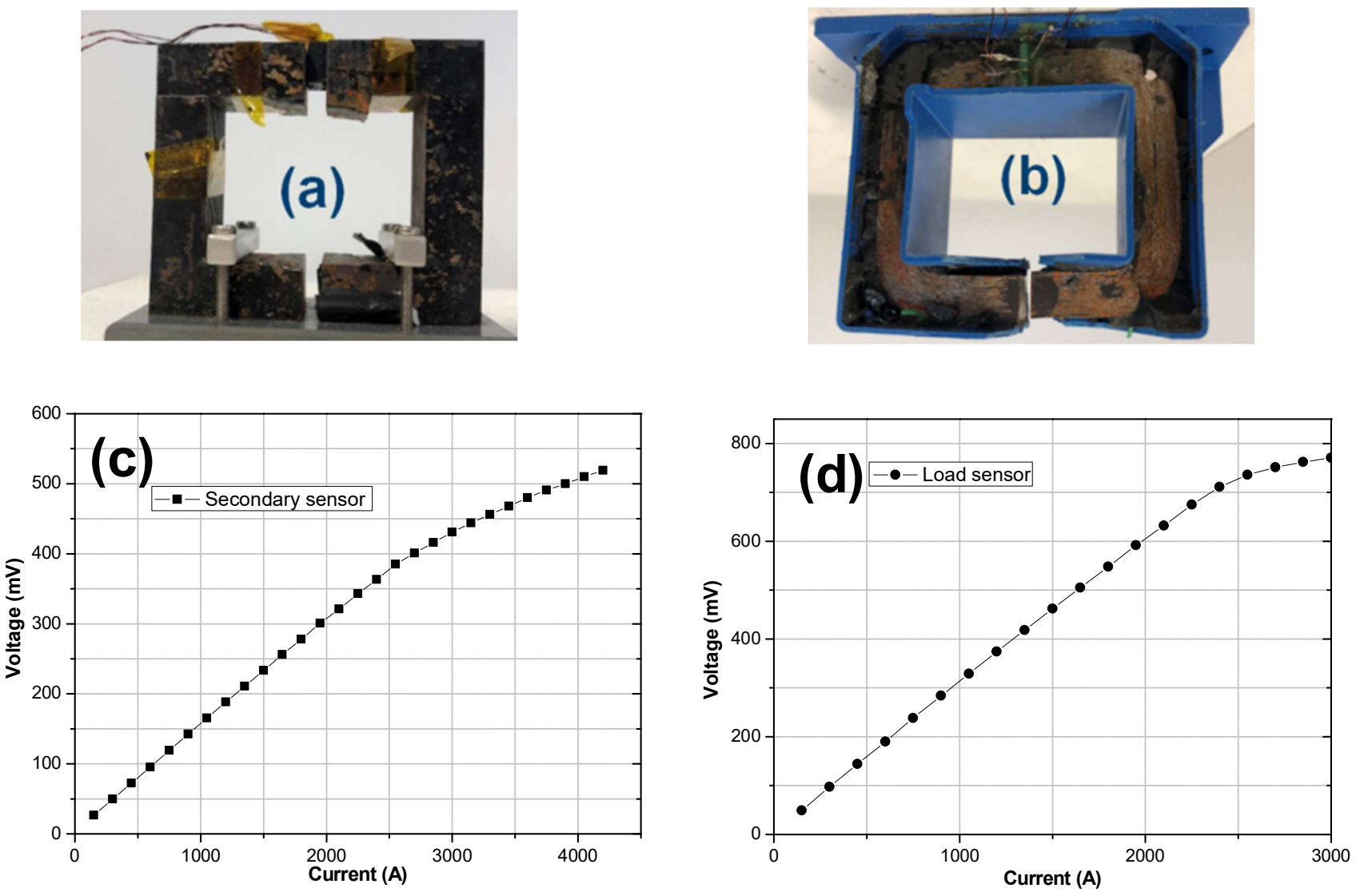

Fig. 2 Pictures of the open loop Hall current sensors and calibration curves. (a) the secondary current sensor, (b) the load current sensor, (c) the calibrated current-Hall voltage curve of the secondary current sensor, (d) the calibrated current-Hall voltage curve of the load current sensor.

Each of the sensors is formed by a magnetic circuit which consists of a pair of 'C' shape cores, and a Hall effect magnetic sensor. In order to measure current up to several thousand Amperes, the total air gap length in the magnetic circuit of each sensor is several millimeters. As can be seen from Fig. 2(c), the secondary current sensor voltage does not saturate until the current exceeds 2800 A. In contrast, the load current sensor output voltage is linear before the current exceeding 2400 A, as shown in Fig. 2(d).

The primary winding of the transformer was powered by a KEPCO-BOP 2020 power amplifier. The KEPCO is able to output a maximum current of 20 A and a maximum voltage of 20 V. The KEPCO was controlled by an NI-PCIe 6343 data acquisition (DAQ) card which can output programmable analogue signal from LabVIEW. The KEPCO worked in current mode, in which the output current is proportional to the input signal. The two switching magnets are powered by an EP4000 audio amp. The audio amp has two output channels. Each channel powered one of the magnets. The EP 4000 was also controlled by a programmed analogue signal from the DAQ. The switching magnetic fields were not recorded. This is because the air gaps of the magnets are not homogeneous so it is difficult to exactly determine the field value. Both the secondary current and the load current signals were acquired by the DAQ card at a sampling rate of 1 kHz. In this paper, during all measurements the superconducting circuit was immerged into liquid nitrogen at 77 K.

## 3 Experiments and results

### 3.1 flux-pumped current over 1 kA

The operation of the flux pump is similar to that described in our previous work [21] [27]. A 1

Hz sine-wave current was applied to the transformer primary to induce the secondary current. The peak value of the secondary current can be over 2900 A before exceeding the critical current of the secondary winding. To maintain a safety margin, we only induced a secondary current with peak-to-peak value of 5 kA during flux pumping. The switch magnets were powered by an intermittent 50 Hz sine-wave current, which was only applied 10 continuous cycles around the positive peak of the secondary current during each cycle of the secondary current. Fig. 3 shows the waveform of secondary current during flux pumping. The current waveform is not perfectly symmetrical, and the positive peak current is lower than the negative peak current. This is because the switching field is only applied around the current positive peak, causing a reduction in the positive current.

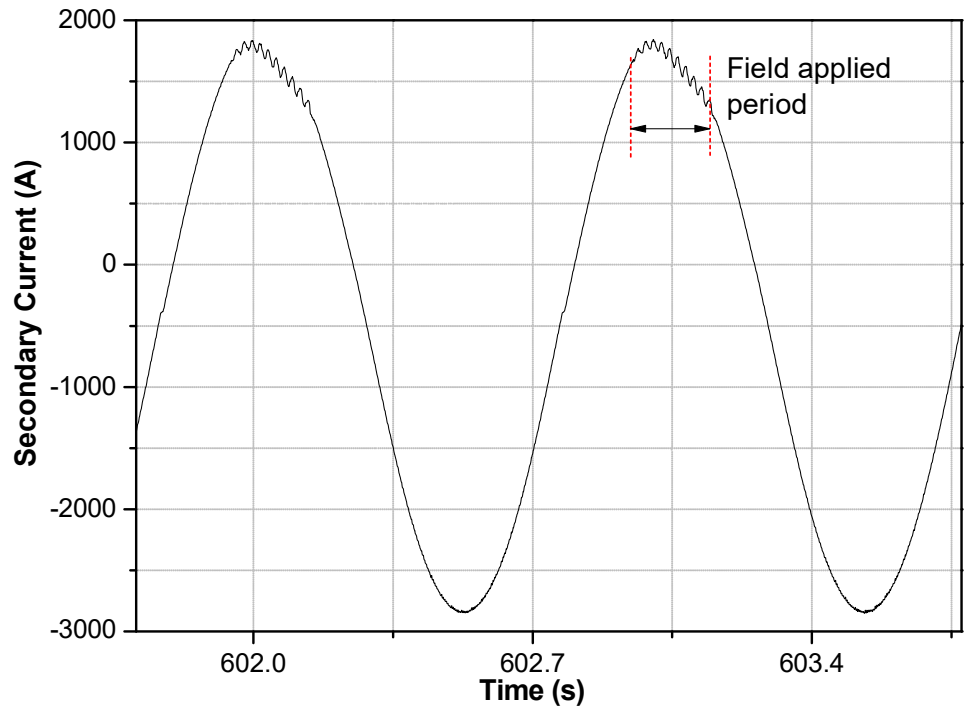

Fig. 3 Waveform of the transformer secondary current. The peak-to-peak value is slightly lower than 5 kA. There is a DC bias due to the switching field which was only applied around the positive peak.

We pursued the highest pumped current the device could output. The power supply for the switching magnets was adjusted to maximum. The load current curve is shown in Fig. 4. Within 200 s time, the load was charged to over 1150 A. To our knowledge, this is the highest direct current ever generated by an HTS flux pump, and it is the only kilo-ampere level pumped current reported to date. According to the trend of the curve, the current should have saturated at a much higher current level. However, prior to saturation the current suddenly dropped to nearly zero, indicating a quench occurred. We did not stop the device, so several continuous quenches were observed, and each time the maximum current before quench was rather constant. Although we have not found the reason for the quench, most probability it is because of exceeding the critical current of the load loop. The three parallel tapes of the load should have a total self-field critical current of over 2 kA. But the critical current of the load loop should be below 2 kA, considering the 4-turn load coil generates a magnetic field which reduces the critical current. Moreover, termination soldering may also incur a reduction of $I_c$. The result indicates that although the device can operate HTS magnets safely a low current, there could be risks in high current operations, which should be taken into consideration when designing higher current flux pumps.

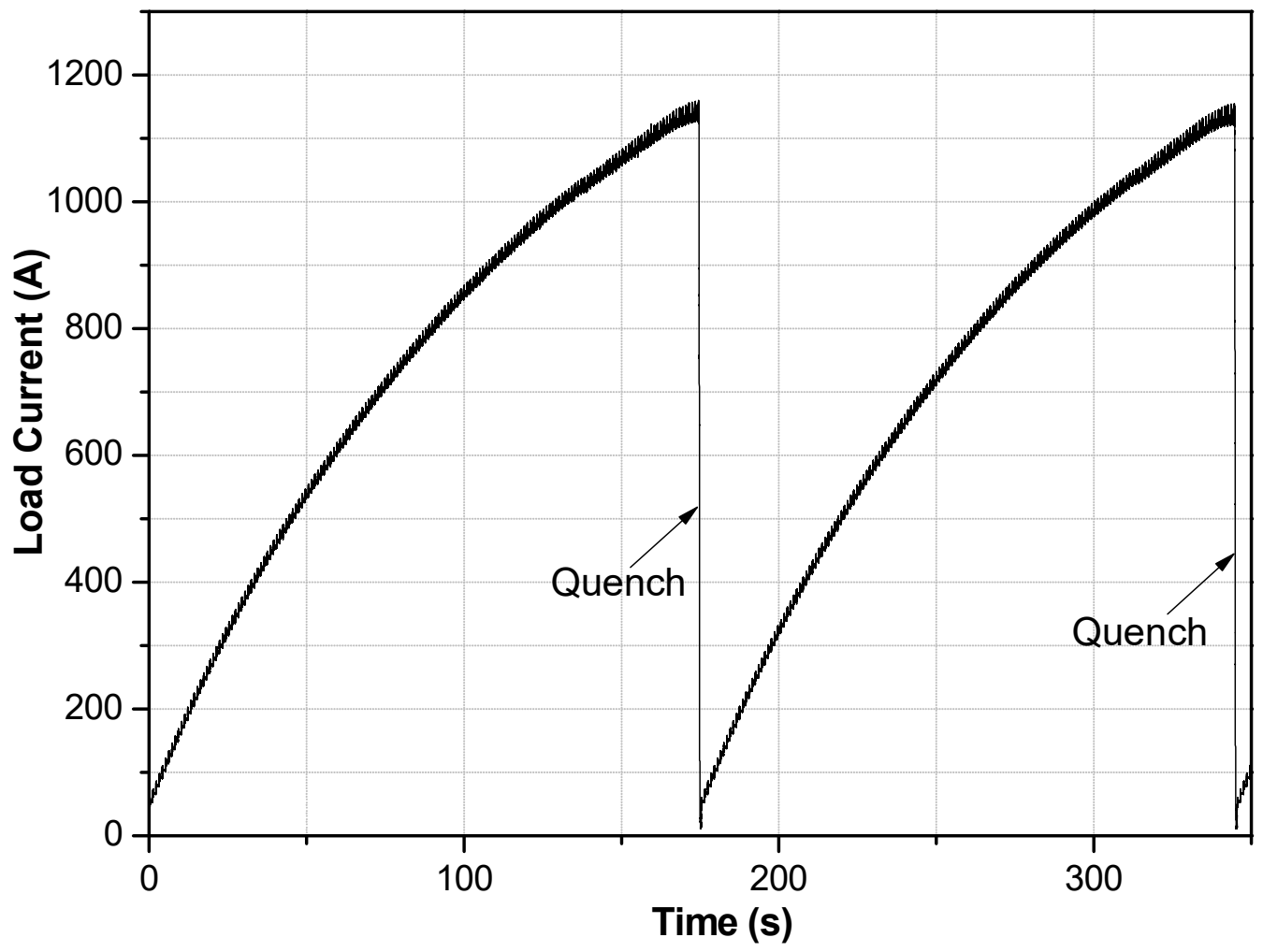

Fig. 4. Charging curve of the flux pump. Quench was observed after the load current exceeded 1150 A.

### 3.2 Field ripple and noises

To maintain a quasi-persistent current operation, we reduced the output current of the device. This was achieved by adjusting the phase angle between the switching field and the transformer secondary current; so that the switching field was misalign with the peak of the secondary current.

The charging curves under two different magnitudes of switching field are shown in Fig. 5. At a higher switching field, the load current reached over 1100 A, and at a lower field the load current saturated at 650A.

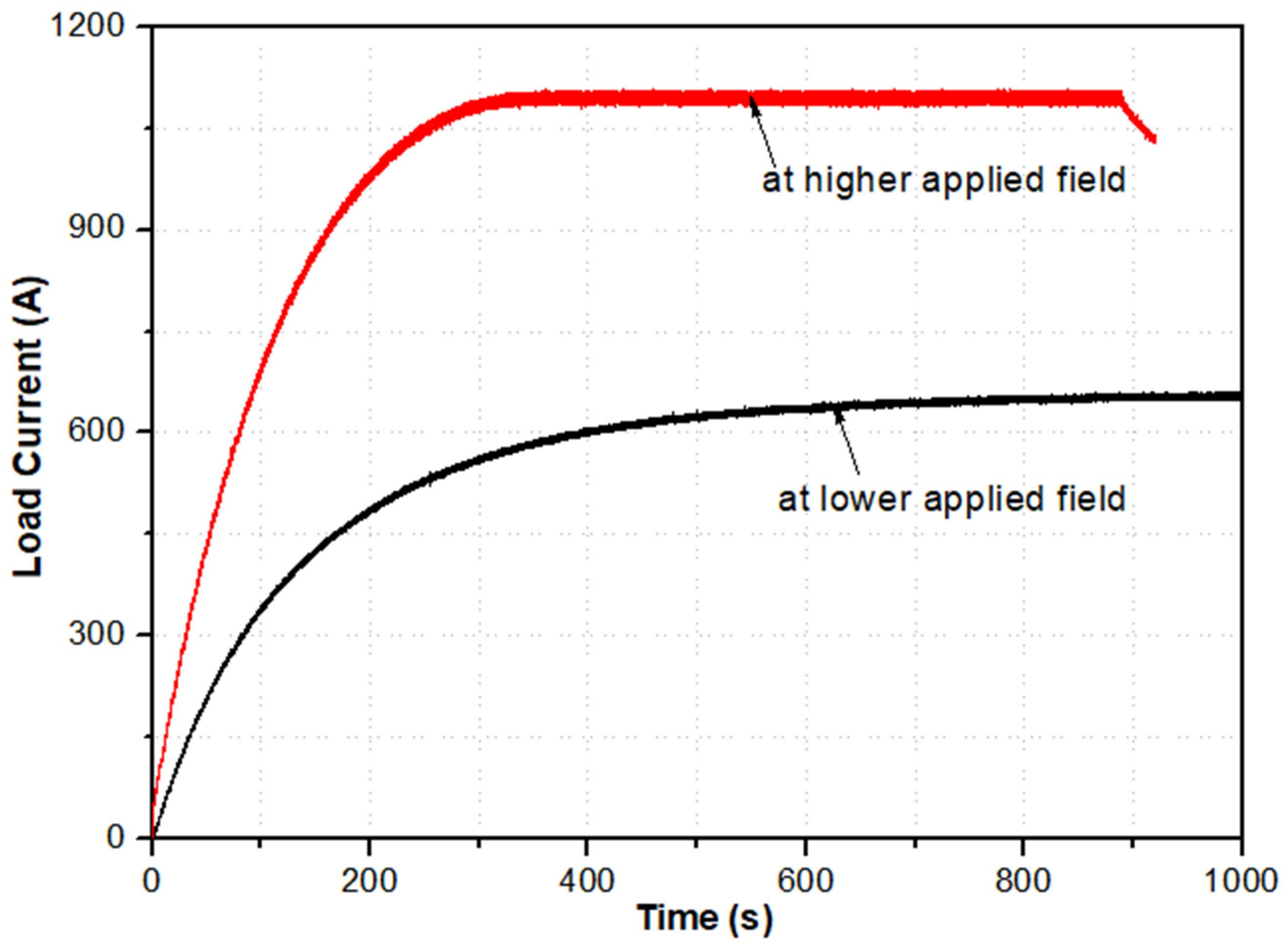

Fig. 5 Curves showing the load current saturated at different levels.

From Fig. 5 it can be seen that the current ripple at 1100A is much higher than that at 650 A. The detail of the ripple is enlarged in Fig. 6. It should be noted that all the data are originally sampled by the DAQ card, without any hardware or software filtering. At 1100 A current level, the maximum load current oscillation is about 20 A, and at 650 A current level, it is about 10 A.

Considering that the load inductance is less than 10 $\mu$H. The flux ripple at 1100 A is less than 0.2 mWb at 1100 A, and it is less than 0.1 mWb at 650A.

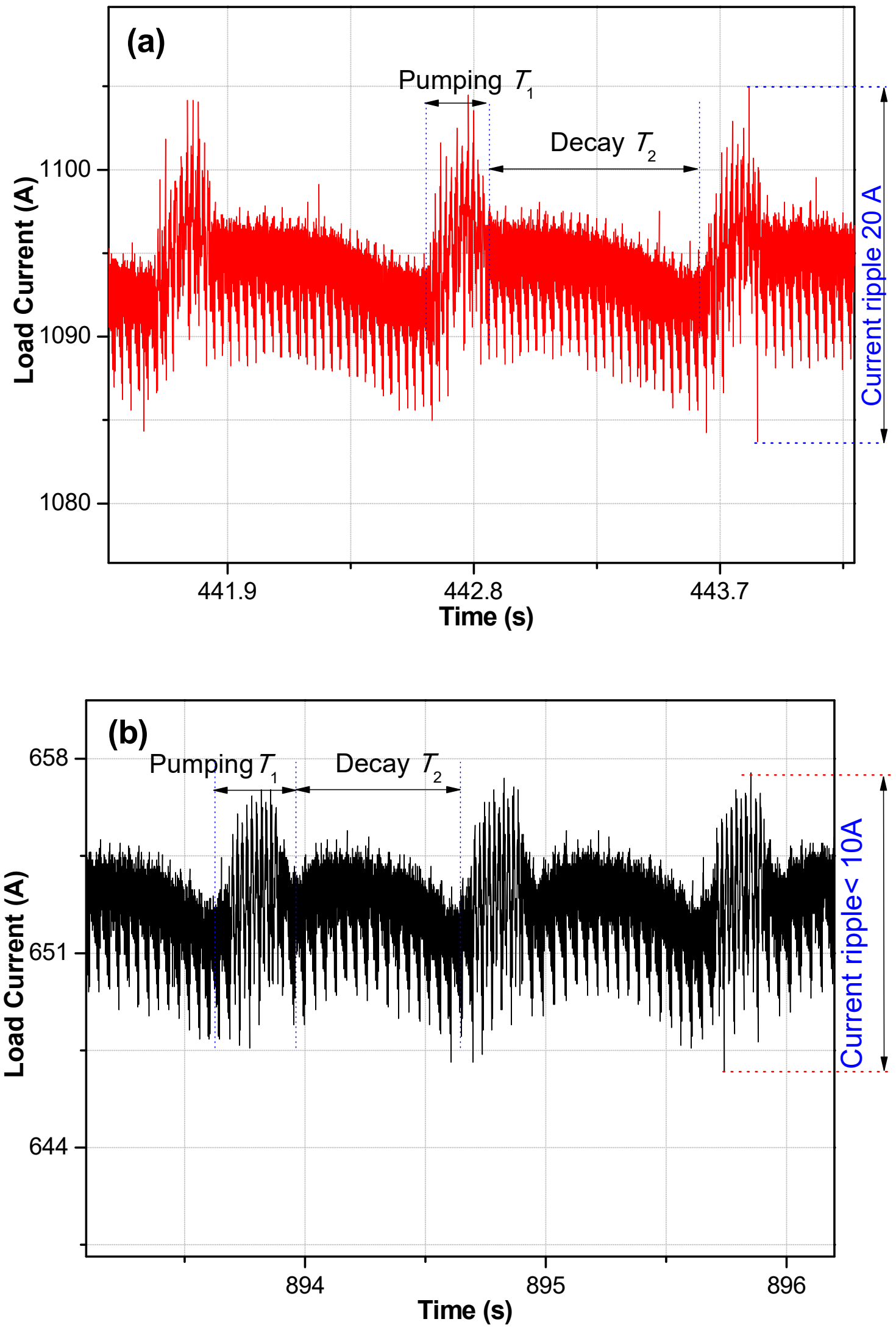

Fig. 6. Details of load current ripple during flux pumping stage and decay stage. (a) load current stabilizes at 1100 A, (b) load current stabilizes at 650 A.

There are several sources of the current ripple.

The first is from the current decay. Within each cycle of the secondary current, there is a flux pumping stage and a decay stage, as shown in Fig. 6. During the decay stage, the total flux leakage can be considered as:

$$\Delta\Phi = R_L \times I_L \times \Delta T_2 \quad (1)$$

This part of magnetic field oscillation is proportional the load current $I_L$, the load loop resistance $R_L$, and the decay time period $\Delta T_2$. The load resistance $R_L$ is mainly contributed by the joint resistance at lower load current level. At higher load current level, it is largely affected by flux creep and external oscillating magnetic fields [28]. To reduce $R_L$, it is desirable to reduce the joint resistance, to operate at a lower load current level, and to shield the load loop from external ripple fields. It is also possible to reduce the load current ripple by reducing the load current $I_L$, or by reducing the decay time $\Delta T_2$ during each cycle (which can be achieved by increasing the secondary current frequency).

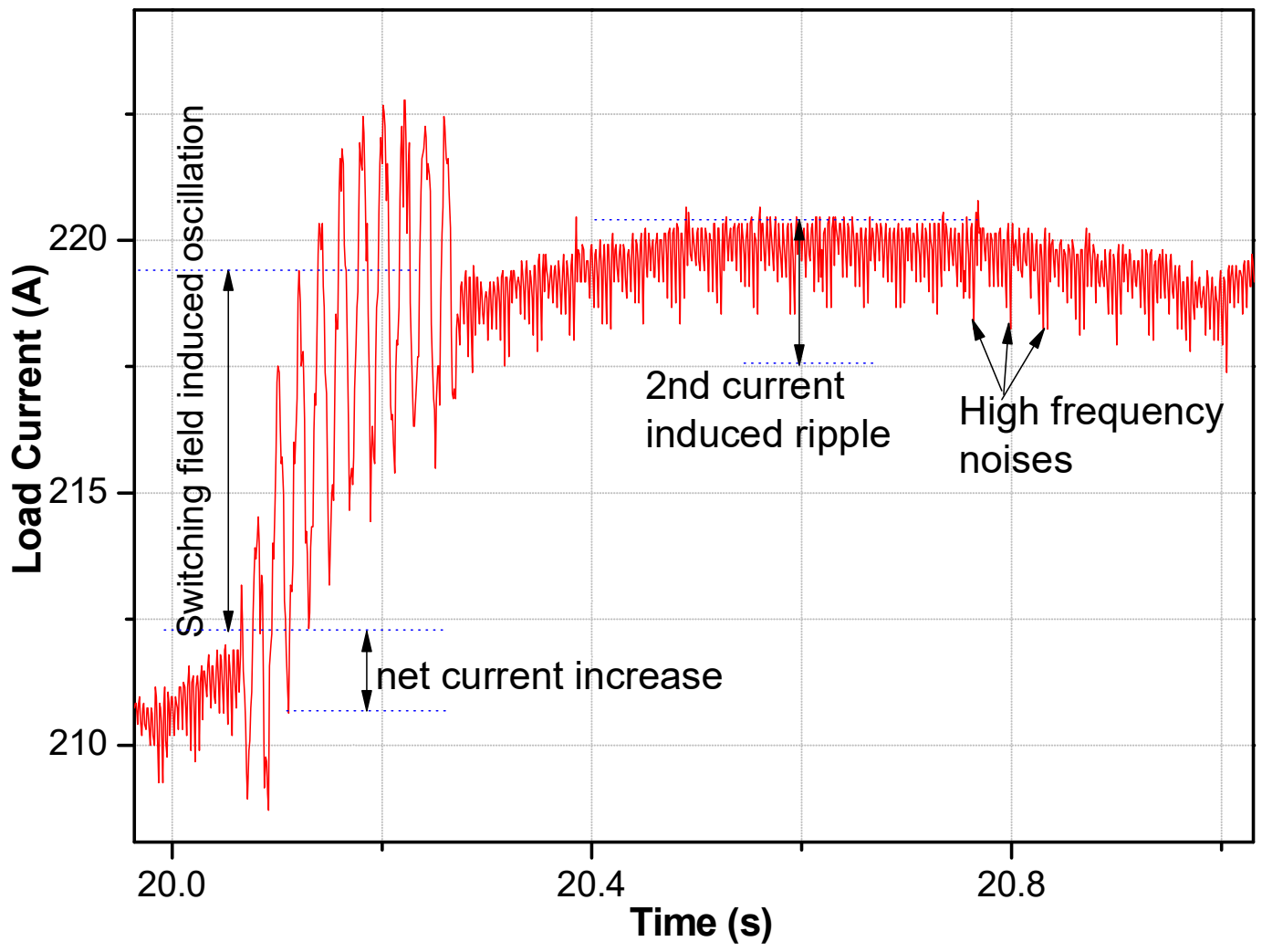

Fig. 7. Details of the load current curve during charging. Apart from a net current increase, there are several kinds of oscillation, including switching field induced oscillation, transformer secondary current induced ripple, and high frequency noises.

The second source of field ripple is from the flux pumping stage. To compensate the above mentioned current decay, the same amount of flux should be pumped into the load during each pumping period $T_1$, as shown in Fig. 6. Ideally this flux increase does not incur oscillation. Practically however, apart from the net increase, there is also flux oscillation in each cycle of applied switching field, as shown in Fig. 7 (For a clearer view, the data during the current ramping-up rather than saturation is shown). This is on one hand due to mutual induction between the switching magnets and the load loop, which has been minimized by using a bifilar structure bridge superconductor; on the other hand, the bi-directional motion of flux in the bridge superconductor also contributes to the oscillation. This bi-directional flux motion is due to the fact that transport current in the bridge superconductor is less than the critical value, so that some flux enters and exits the bridge superconductor from the same edge [17], and it is associated with a magnetization loss [29]. To reduce the switching field induced oscillation, the possible methods include designing better decoupled circuit, reducing the applied field magnitude, and increasing the bridge current when the switching field is applied.

The third source of field ripple comes from the mutual induction between the transformer secondary loop and the load loop. Although we used a bifilar structure to minimize the mutual inductance, the mutual induction cannot be totally eliminated. The secondary current induced ripple in the load current has been shown in Fig. 7. Apart from the ripple, the high secondary alternating current also results in a current decay. To investigate these effects, we performed two experiments. The first is the free decay. After charging the load to over 1100 A, we switched off the power supplies, so there was no output from either the transformer or the switch magnets. The second is decay with the secondary current on. In this case after charging the load to over 1100 A, we turned off the magnetic switches' power supply but leave the transformer secondary current on.

Fig. 8 shows the current decay curves under these two circumstances. It is clear that the free decay rate is slower than the decay rate with the secondary current on, which proves that the secondary current oscillation incurs loss in the secondary circuit.

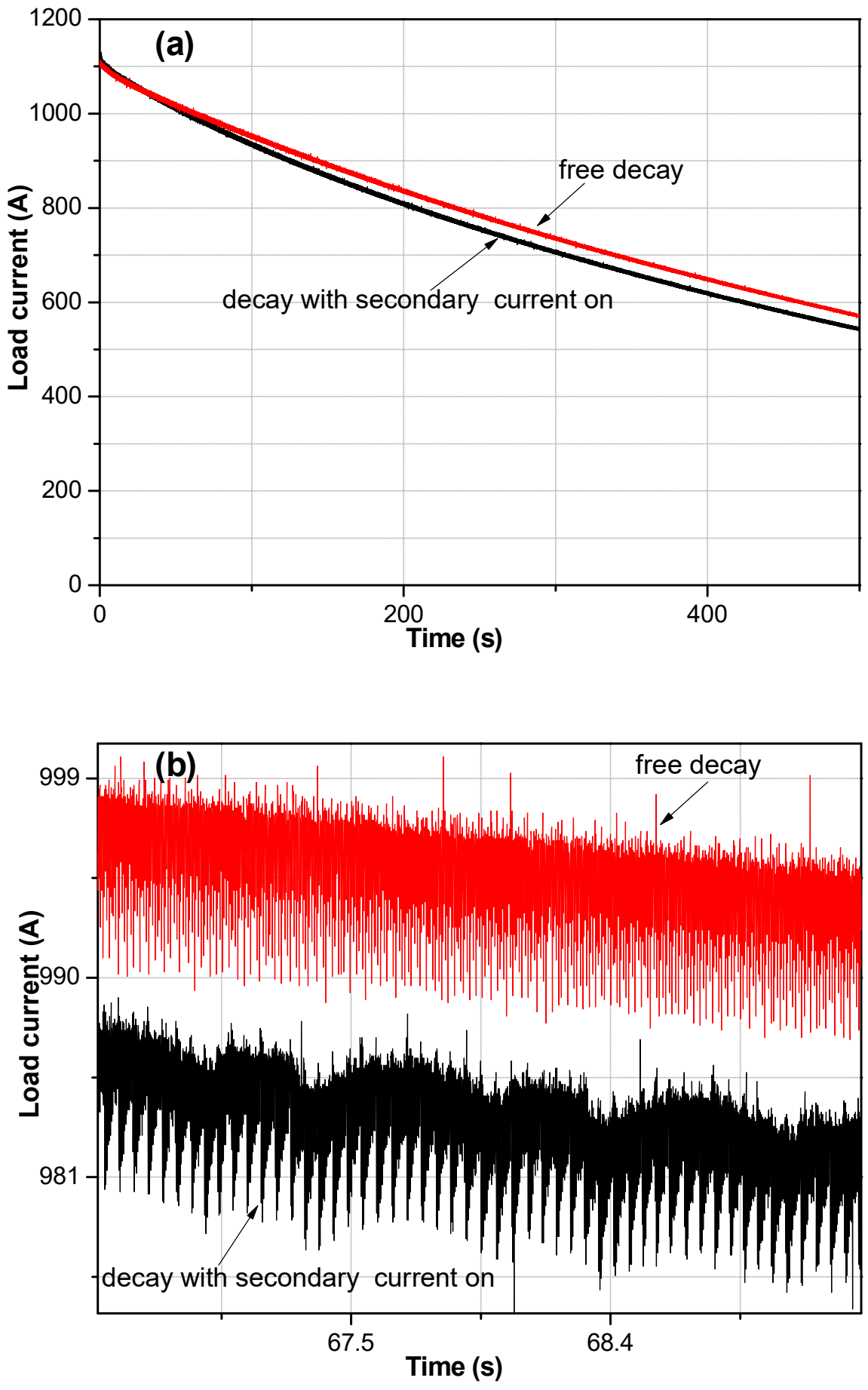

Fig. 8. Details of load current decay. Free decay and decay with secondary current were compared.

In addition to the above analyzed sources of load current ripple, there are also noises coupled to the measurements. The noises can be seen from Fig. 7 and Fig. 8(b). These noises come from the power supplies, the sensors themselves, and the DAQ errors. From Fig. 8(b) we can see that in the free decay curve, the noise is mainly high frequency components; whereas for the curves with secondary current-on, there is not only a one-Hertz component which is the induction from the transformer secondary, but also some higher frequency harmonics which may come from the KEPCO power supply.

Despite the fact that there are various sources of field ripple and noise in the load, the raw experimental data show that the total flux fluctuation is under 0.2 mWb at 1100 A and 0.1 mWb at 650 A. It should be noted that from circuit point of view, the flux pump is a voltage source rather than a current source, therefore the flux ripple can be considered constant irrespective of inductance value. In this case the flux fluctuation has already been satisfactory for the temporal field stability of practical sized NMR/NMR magnets.

## 4 Summary

In this paper, we developed and tested a high-current transformer-rectifier HTS flux pump prototype. The device managed to output a quasi-persistent current of over 1 kA, which is the highest flux pumped current in an HTS magnet ever reported. In the prototype, we exemplified the feasibility of using parallel high current tapes as the transformer secondary, and parallel coils as the load to achieve high current. The result also shows that the dynamic resistance switched transformer-rectifier flux pump is scalable, and could be an ideal candidate for generating ultra-high current for compact high field HTS magnets. Moreover, at high load current level the device can maintain satisfactory field stability. The raw data have shown that at 1 kA current level, the total flux ripple in the load is less than 0.2 mWb. Although this field stability has already been satisfactory for the 1ppm requirement of MRI/NMR, it could be substantially improved if we properly design and manufacture the circuit and develop effective feedback control algorithms.

## Acknowledgements

This work is supported by EPSRC Grant No. EP/R016615/1. The authors would like to thank J Grundy for his help in setting up the experiment.

## References

[1] H. Weijers, et al, “Progress in the Development and Construction of a 32-T Superconducting Magnet,” IEEE Transactions on Applied Superconductivity, Vol. 26, No. 4, 2016.

[2] T. A. Painter, et al “An Integrated Coil Form Test Coil Design for High Current REBCO DC Solenoids”, submitted to IEEE Transactions on Applied Superconductivity.

[3] C. Hoffmann, D. Pooke, and A. D. Caplin, “Flux pump for HTS magnets,”IEEE Trans. Appl. Supercond., vol. 21, no. 3, pp. 1628–1631,Jun. 2011.

[4] Z. Bai *et al.*, “A newly developed pulse-type microampere magnetic flux pump,”*IEEE Trans. Appl. Supercond.*, vol. 20, no. 3, pp. 1667–1670, Jun. 2010.

[5] Z. Bai, G. Yan, C. Wu, S. Ding, and C. Chen, “A novel high temperature superconducting magnetic flux pump for MRI magnets,” *Cryogenic*, vol. 50 , no. 10, pp. 688–692, Oct. 2010.

[6] L. Fu, K. Matsuda, and T. A. Coombs, “Linear Flux Pump Device Applied to HTS Magnets: Further Characteristics on Wave Profile, Number of Poles, and Control of Saturated Current,” IEEE Trans. Appl. Supercond., vol. 26, no. 3, Apr., 2006, Art. ID 0500304.

[7] R. M. Walsh, R. Slade, P. Donald, and C. Hoffmann, “Characterization of current stability in an HTS NMR system energized by an HTS flux pump,” IEEE Trans. Appl. Supercond., vol. 24, no. 3, Jun. 2014,Art. ID 4600805.

[8] Z. Jiang et al., “Impact of flux gap upon dynamic resistance of a rotating HTS flux pump,”Supercond. Sci. Technol., vol. 28, no. 11, Nov. 2015, Art. ID 115008.

[9] S. Lee et al., “Persistent Current Mode Operation of A2G HTS Coil With A Flux Pump,” IEEE Trans. Appl. Supercond., vol. 26, no. 4, Jun. 2016, Art. ID 0606104.

[10] Z. Jiang, C. W. Bumby, R. A. Badcock, and H.-J. Sung, “A novel rotating HTS flux pump incorporating a ferromagnetic circuit,” IEEE Trans Appl. Supercond.,vol. 26, no. 2, Mar. 2016, Art. ID 4900706.

[11] C. W. Bumby et al., “Through-Wall Excitation of a Magnet Coil by an External-Rotor HTS Flux Pump,” IEEE Trans. Appl. Supercond., vol. 26, no. 4, Jun. 2016, Art. ID 0500505.

[12] Y. H. Choi, et al, “A Study on Charge–Discharge Characteristics of No-Insulation GdBCO Magnets Energized via a Flux Injector”, IEEE Trans. Appl. Supercond., vol. 27, no. 4, Jun. 2017, Art. ID 4601206.

[13] H. Jeon, et al, "PID Control of an Electromagnet-Based Rotary HTS Flux Pump for Maintaining Constant Field in HTS Synchronous Motors", IEEE Trans. Appl. Supercond., vol. 28, no. 4, Jun. 2018, Art. ID 5207605.

[14] W. Wang, Y. Lei, S. Huang, P. Wang, Z. Huang, Q. Zhou, "Charging 2G HTS Double Pancake Coils With a Wireless Superconducting DC P ower Supply for Persistent Current Operation" , IEEE Trans. Appl. Supercond., vol. 28, no. 3, Apr. 2018, Art. ID 0600804.

[15] Z. Jiang, K. Hamilton, N. Amemiya, R. A. Badcock, and C. W. Bumby, "Dynamic resistance of a high-Tc superconducting flux pump,"Appl. Phys. Lett., vol. 105, no. 11, Sep. 2014, Art. ID 112601.

[16] Andrianov V V, Zenkevich V B, Kurguzov V V, Sychev V V and Ternovskii F F 1970 Sov. Phys. JETP 31 815.

[17] Oomen M P, Rieger J, Leghissa M, ten Haken B and ten Kate H H J 1999 Supercond. Sci. Technol. 12 382.

[18] Jiang Z, Toyomoto R, Amemiya N, Zhang X and Bumby C W, 2017 Supercond. Sci. Technol. 30 03LT01.

[19] C. W. Bumby, Z. Jiang, A. E. Pantoja, and R. A. Badcock, "Anomalous open-circuit voltage from a high-Tc superconducting dynamo," Appl. Phys. Lett., vol. 108, no. 12, Mar. 2016, Art. ID 122601.

[20] J. Geng et al., "Origin of dc voltage in type II superconducting flux pumps: field, field rate of change, and current density dependence of resistivity," J. Phys. D: Appl. Phys., vol. 49, no. 11, Feb. 2016, Art. no. 11LT01.

[21] J. Geng and T. A. Coombs, "Mechanism of a high-Tc superconducting flux pump: Using alternating magnetic field to trigger flux flow," Appl. Phys. Lett., vol. 107, no. 14, Oct. 2015, Art. no. 142601.

[22] J. Geng and T. A. Coombs, "An HTS flux pump operated by directly driving a superconductor into flux flow region in the E–Jcurve," Supercond. Sci. Technol., vol. 29, no. 9, Jul. 2016, Art. no. 095004.

[23] A. M. Campbell, Supercond. Sci. Technol. 30, 125015 (2017).

[24] J. Geng and T. A. Coombs, Supercond. Sci. Technol. 31, 125015 (2018).

[25] K. Hamilton, et al, "Design and Performance of a "Squirrel-Cage" Dynamo-Type HTS Flux Pump", IEEE Trans. Appl. Supercond., vol. 28, no. 4, Jun. 2018, Art. ID 5205705.

[26] L. Fu, K. Matsuda, T. Lecrevisse, Y. Iwasa, and T. Coombs, "Linear flux pump device applied to high temperature superconducting (HTS) magnets," Supercond. Sci. Technol., vol. 29, no. 4, Feb. 2016, Art. no. 04LT01.

[27] J. Geng et al., "Operational research on a high-Tc rectifier-type superconducting flux pump," Supercond. Sci. Technol., vol. 29, no. 3, Feb. 2016, Art. no. 035015.

[28] J Geng, H Zhang, C Li, X Zhang, B Shen, TA Coombs, "Angular dependence of direct current decay in a closed YBCO double-pancake coil under external AC magnetic field and reduction by magnetic shielding", Supercond. Sci. Technol., vol. 30, no. 3, 2016, Art. no. 035022.

[29] Martin N. Wilson, Superconducting magnets. Oxford University Press, 2002.